\documentclass[preprint,aps,nofootinbib,prd,showpacs]{revtex4}
\usepackage{graphicx}
\usepackage{amsmath}
\usepackage{amsfonts}
\usepackage{amssymb}
\usepackage{color}%
\usepackage{slashed}

\providecommand{\U}[1]{\protect\rule{.1in}{.1in}}

\newcommand{\f}{\begin{equation}}
\newcommand{\ff}{\end{equation}}
\newcommand{\fa}{\begin{eqnarray}}
\newcommand{\ffa}{\end{eqnarray}}

\begin{document}

\title{Scalar Boundary Conditions in Hyperscaling Violating Geometry}
\author{Jian-Pin Wu $^{1,3}$}
\email{jianpinwu@mail.bnu.edu.cn}
\author{Xiao-Mei Kuang $^{2}$}
\email{xmeikuang@gmail.com}

\affiliation{$^1$ Institute of Gravitation and Cosmology, Department of Physics, School of Mathematics and Physics, Bohai University, Jinzhou 121013, China
\\ $^2$ Instituto de F\'isica, Pontificia Universidad Cat\'olica de Valpara\'iso, Casilla 4059, Valpara\'iso, Chile
\\ $^3$ State Key Laboratory of Theoretical Physics, Institute of Theoretical Physics, Chinese Academy of Sciences, Beijing 100190, China}

\begin{abstract}
We study the possible boundary conditions of scalar field modes in a hyperscaling violation(HV) geometry
with Lifshitz dynamical exponent $z (z\geqslant1)$ and hyperscaling violation exponent $\theta (\theta\neq0)$.
For the case with $\theta>0$, we show that in the parameter range with $1\leq z\leq 2,~-z+d-1<\theta\leq (d-1)(z-1)$ or $z>2,~-z+d-1<\theta\leq d-1$,
the boundary conditions have different types, including the Neumann, Dirichlet and Robin conditions,
while in the range with $\theta\leq-z+d-1$, only Dirichlet type condition can be set.
In particular, we further confirm that the mass of the scalar field does not play any role in determining the possible boundary conditions for $\theta>0$,
which has been addressed in Ref. \cite{1201.1905}.
Meanwhile, we also do the parallel investigation in the case with $\theta<0$. We find that for $m^2<0$, three types of boundary conditions are available, but for $m^2>0$, only one type is available.
\end{abstract}

\maketitle

\section{Introduction}
The use of holographic duality into study of strongly-coupled field theories \cite{hep-th/9711200,hep-th/9905111} has produced
substantial progress in reproducing and understanding phenomena not only from relativistic systems like QCD \cite{1101.0618},
but also from the non-relativistic strongly interacting condensed matter systems \cite{0903.3246,0909.0518,1002.2947,1108.1197}.
These applications have provoked interest in holographic realization of symmetry groups that go beyond of relativistic conformal symmetry.
These include in particular the Schr\"odinger  symmetry\cite{0804.4053,0804.3972},
Lifshitz symmetry\cite{0808.1725,0812.0530,0812.5088,0905.1136,0905.3183,0905.2678,0911.2777,0909.0263,0909.1347,0909.2807,0910.4428,1102.0578,1204.0360,1105.6335}
and HV symmetry \cite{1112.0573,1201.1905,1112.2702,1209.3946,1411.0312,1501.05318},
which exhibit in common the anisotropic scaling characterized by the dynamic critical exponent $z> 1$ between time and space coordinates on the boundary.
Besides, lots of extensive holographic study on non-relativistic scaling geometry have been present
in \cite{Charmousis:2010,1112.5074,1201.1764,1201.3832,JPWu1,JPWu2,JPWu3,1411.3331,1401.7993,1409.2945,1411.5627,1511.03008,1408.6380,1408.0795} and references therein.

In the application of AdS/CFT correspondence, to gain reasonable interpretations of the dual field theory,
it is necessary and important to study possible boundary conditions of various fields in the bulk theory.
In \cite{gr-qc/0305012}, the authors proposed a unique prescription for finding consistent boundary conditions of the fields
 on a stably causal, static spacetime which possesses ``good dynamic". Later,
 by applying this proposal into the scalar field(also electromagnetic and gravitational perturbations)
 in the global anti-de Sitter space in \cite{hep-th/0402184}, they found that the boundary conditions ranging from
 ranging from Dirichlet to Robin to Neumann conditions are all possible depending on the effective mass of the scalar field
 . Recently, inspired by \cite{gr-qc/0305012,hep-th/0402184},
this method of disclosing the possible boundary conditions is extended into the study of the scalar field
in Lifshitz geometry in \cite{1212.1728,1212.2572}.
It was addressed that depending on the effective mass of the scalar field, which is determined by the Lifshtiz exponent and the dimension of spacetime,
the scalar field can have no reasonable dynamic, one-parameter choice of boundary conditions type and only one boundary condition type.

In this letter, we shall explore the possible boundary conditions of scalar field in the HV geometry closely
following the approach of \cite{1212.1728}. Using the Poincar$\acute{e}$ patch, the HV geometry has the following form\cite{Charmousis:2010}
\begin{eqnarray}\label{HyperscalngGeometry}
ds_{d+1}^2=u^{\frac{2\theta}{d-1}}\left(-\frac{1}{u^{2z}}dt^2+\frac{1}{u^2}du^2+\frac{1}{u^2}d\vec{x}^2\right),
\end{eqnarray}
where $x$ runs over $d-1$ dimensions and the boundary of this geometry is at $u=0$.
The above metric enjoys both a Lifshitz dynamical critical exponent $z$ ($z\geq 1$) and a HV exponent $\theta$.
Under the scale-transformation
\begin{equation}
t\rightarrow\lambda^zt,\,\, x_i\rightarrow\lambda x_i,\,\,
u\rightarrow\lambda u~, \label{scaling}
\end{equation}
the metric (\ref{HyperscalngGeometry}) transforms as $ds\rightarrow\lambda^{\theta/(d-1)}ds$, which breaks the scale-invariance.

Before proceeding, we shall give several comments on the HV geometry (\ref{HyperscalngGeometry}).
Firstly, the scalar curvature is $R\propto u^{-\frac{2\theta}{d-1}}$, so that
there is a curvature singularity at $u=0$ for $\theta> 0$\cite{1201.1905,1408.0795}.
However, to implement the $\theta\neq 0$ HV geometry (\ref{HyperscalngGeometry}),
one usually needs to introduce a dilaton field (see for example \cite{1201.1905,1209.3946,1411.0312}),
which can be used to absorb the curvature singularity into the dilaton in the dual frame.
Furthermore, the null energy condition gives us the following constrains on  $z$ and $\theta$,
\begin{eqnarray}
&&
(d-1-\theta)((d-1)(z-1)-\theta)\geq 0~,
\
\\
&&
(d-1-\theta+z)(z-1)\geq 0~.
\end{eqnarray}
At the same time, the stabilities in the gravity side require $\theta\leq d-1$.
Combining the above discussions, we summarize the allowed $z$ and $\theta$  as
\begin{eqnarray}
&&
1\leq z\leq 2,~~~~\theta\leq (d-1)(z-1)~,
\label{NEC1}
\\
&&
z>2,~~~~\theta\leq d-1~.
\label{NEC2}
\end{eqnarray}

In section II, we present our detailed analysis on the possible conditions of the scalar field with the hyperscaling violation exponent $\theta>0$
and $\theta<0$, respectively. Section III is the conclusion and discussion. In Appendix A, we give a brief summary for the series solutions of
the linear differential equation, which we use frequently in the main context.
In addition, the discussions on the square integrability of solutions at infinity are presented in Appendix B.

\section{Scalar boundary conditions in HV geometry}\label{HV}
Since the HV geometry (\ref{HyperscalngGeometry}) we will study is static and stably causal,
it has well defined timelike hypersurface and each is equivalent to any other.
Therefore, we can explore the possible boundary conditions for scalar field $\Psi$ on
a particular hypersurface $\Sigma$.
After selecting a static direction $t$, the Klein-Gordon (KG) equation can be written as
\begin{eqnarray}\label{KGv1}
-u^{2z-\frac{2}{d-1}\theta}\partial_t^2\Psi + u^{d+z-\frac{d+1}{d-1}\theta}\partial_u(u^{-z-d+2+\theta}\partial_u\Psi)+\nabla^i\nabla_i\Psi-m^2\Psi=0.
\end{eqnarray}
We are interesting in momentum space and expand $\Psi$ as
\begin{eqnarray}\label{phiMomentum}
\Psi=u^p\int d\vec{k}c_{\vec{k}}e^{i\vec{k}.\vec{x}}\psi_k(t,u),~~~\mathrm{with}~~~p=\frac{z-\theta+d-2}{2}.
\end{eqnarray}
Under the above transformation, the KG equation has the form
\begin{eqnarray}\label{KGMomentum}
\partial_t^2\psi_k=-A\psi_k,
\end{eqnarray}
where we have defined an operator $A$ as
\begin{eqnarray}\label{A}
A=-u^{2(1-z)}[\partial_u^2-p(p+1)u^{-2}-m^2u^{-2+\frac{2}{d-1}\theta}-k^2].
\end{eqnarray}
It can act on the Hilbert space $\mathcal{H}=L^2(\Sigma,\mu)$ of functions living on the time slice and being square integrable
with the measure $\mu=V^{-1}d\Sigma$.

For the HV geometry, the measure is
\begin{eqnarray}\label{Measure}
\mu=u^{\theta+z-d}du d\vec{x}_{d-1}.
\end{eqnarray}
So, the inner product on the Hilbert space $\mathcal{H}$ can be calculated as
\begin{eqnarray}\label{InnerProduct}
\langle\Psi_2|\Psi_1\rangle=\int_{\Sigma}\Psi_2^\ast\Psi_1 u^{\theta+z-d}dud\vec{x}_{d-1}.
\end{eqnarray}
Obviously, the set of functions included in the Hilbert space not only depend on Lifshitz exponent $z$ but also $\theta$.
Assuming that the wave packets are normalizable in the $\vec{x}$ directions, the inner product (\ref{InnerProduct}) becomes in term of $\psi_k(t,u)$
\begin{eqnarray}\label{InnerProductMomentum}
\langle\Psi_2|\Psi_1\rangle_u=\int du u^{2z-2}\psi_{2,k}^\ast\psi_{1,k}.
\end{eqnarray}
Subsequently, we shall study possible boundary conditions at $u=0$ for
$\psi_k(t,u)$, which satisfy the K-G equation and are finite under the norm (\ref{InnerProductMomentum}),
for the cases with $\theta>0$ and $\theta<0$, respectively.
The related study for $\theta=0$ can be found in \cite{1212.1728,1212.2572}.

\subsection{Boundary conditions for $\theta>0$}\label{theta1}

\subsubsection{Solution spaces $\mathcal{K}_{\pm}$}
In order to classify the possible boundary conditions, we only need to know the solution of modes near the boundary and its normalizability.
Therefore, we shall only explore the solution spaces $\mathcal{K}_{\pm}$ at $u=0$ in the main body of this letter\footnote{For completeness,
we shall also present the discussions on the square integrability of solution spaces $\mathcal{K}_{\pm}$ over the full radial range from $u=0$ to $u=\infty$ in Appendix B.},
whose elements satisfy the following eigenequation\footnote{For convenience, we will drop the subscript $k$ in $\psi$.}
\begin{eqnarray}\label{APsi}
A\psi=\lambda\psi,
\end{eqnarray}
with eigenvalues $\lambda={\pm}i$ and $A$ defined in Eq.(\ref{A}).
In addition, the elements of $\mathcal{K}_{\pm}$ also should be square integrable under the measure (\ref{Measure}).

Substituting the expression of $A$, the eigenequation (\ref{APsi}) can be rewritten as
\begin{eqnarray}\label{KGv2}
\partial_u^2\psi+Q(u)\psi=0,
\end{eqnarray}
where
\begin{eqnarray}\label{Qu}
Q(u)=-p(p+1)u^{-2}-m^2u^{\frac{2}{d-1}\theta-2}-k^2+\lambda u^{2z-2}.
\end{eqnarray}
Note that when $\theta=0$, Eq.(\ref{KGv2}) recovers to the equation studied in the Lifshitz-AdS geometry\cite{1212.1728,1212.2572}.
The above eigenequation is a second order linear differential equation and so there exist two linearly independent solutions
for any given eigenvalue $\lambda$. However, it is difficult to find exact analytic solutions for Eq.(\ref{KGv2}) as that happens in AdS geometry \cite{hep-th/0402184}.
Fortunately, because what we are interesting in is the normalizability of
the solutions for eigenequation (\ref{KGv2}) under the measure (\ref{InnerProductMomentum}),
we only need to explore the behavior of eigenfunction $\psi_{\pm}$ near the boundary regardless of their full expression.

Since for $\theta>0$,
\begin{eqnarray}
\label{q00}
q_0=\lim_{u \rightarrow 0}u^2Q(u)=-p(p+1),
\end{eqnarray}
which is finite in the limit of $u \rightarrow 0$,
according to the description in Appendix A, we can judge that the singular point $u=0$ is a regular singular point.
Thus, we obtain the following initial equation
\begin{eqnarray}
\label{Indicial0}
\alpha(\alpha-1)-p(p+1)=0,
\end{eqnarray}
which gives two values
\begin{eqnarray}
\label{alpha12}
\alpha_{1}=\frac{1}{2}-\nu,~~~\alpha_{2}=\frac{1}{2}+\nu,
\end{eqnarray}
with
\begin{eqnarray}
\label{nu}
\nu=\sqrt{\frac{1+4p(p+1)}{4}}=\sqrt{\left(\frac{z-\theta+d-1}{2}\right)^2}.
\end{eqnarray}
From the above equation,  it is obvious that $\nu^2$ is independent of the mass of scalar field,
which is very different from the case in AdS geometry or Lifshitz-AdS geometry\footnote{Similar case has been found
in Dirac equation in \cite{1209.3946,1409.2945,1411.5627}.}. This observation agrees well with that mass term does not
contributes to the UV behavior of Green function of scalar operator as addressed in \cite{1201.1905}. 
$\nu^2\geq 0$ is required to ensure the positive definiteness of the operator $A$ on Hilbert space.
Therefore, we have $z-\theta+d-1\geq 0$, i.e., $\theta\leq z+d-1$ which gives us from Eq.(\ref{nu}) that $\nu=\frac{z-\theta+d-1}{2}$.
Note that the case with $\theta> z+d-1$ is also excluded due to the instabilities of gravity according to Eq.(\ref{NEC1}) and Eq.(\ref{NEC2}).

Now we are ready to extract our solutions from Eqs. (\ref{Frobeniusv11})-(\ref{Frobeniusv3}) in the Appendix.
For clarification, we list the leading behaviors of $\psi$ near $u=0$ depending on the values of $\nu$:
\begin{description}
  \item[a.] When $\nu$ is neither zero nor a half integer, we have the leading behaviors
\begin{eqnarray}\label{alpha12}
\psi_{\downarrow}=a_0 u^{\alpha_1},~~~\psi_{\uparrow}=b_0 u^{\alpha_2}.
\end{eqnarray}
  \item[b.] %
When $\nu$ vanishes, i.e., $\alpha_1=\alpha_2=1/2$, to the leading order, the asymptotic behaviors at $u=0$ become
\begin{eqnarray}
\label{alpha1=alpha2}
\psi_{\downarrow}=a_0 u^{1/2},~~~\psi_{\uparrow}=a_0 u^{1/2}\ln u.
\end{eqnarray}
  \item[c.] When $\nu$ becomes some nonzero half integer, i.e., $z-\theta+d-1=j,~j=1,2,\ldots$,
the logarithmic behavior in Eq.(\ref{Frobeniusv3}) diverges.
To have well-defined behavior of $\psi$, the constant $C$ in Eq.(\ref{Frobeniusv3}) should vanish.
So in this case, the asymptotic behaviors are the same as Eq.(\ref{alpha12}).
\end{description}

After having fixed the asymptotic behaviors at $u=0$, we shall study their square integrability under the measure (\ref{InnerProductMomentum}),
in which the leading behavior of $u^{2z-2}\psi^2$ is required in the form of $u^\beta$ with $\beta>-1$.
We first discuss the case of $\nu$ being non-zero, which has the asymptotic behavior as (\ref{alpha12}).
For the solution $\psi_{\uparrow}$, we have
\begin{eqnarray}
u^\beta=u^{2z-2+2\nu+1},
\end{eqnarray}
which results in $\psi_{\uparrow}$ being square integrable when $\nu>-z$.
Recalling the condition of gravitational stability and the expression $\nu=\frac{z-\theta+d-1}{2}>0$, we have the conditions $z-\theta+d-1\geq 0$ and
 $\nu>-z$, which are always satisfied.
Thus, the solution $\psi_{\uparrow}$ is always square integrable near $u=0$.
While $\psi_{\downarrow}$ leads to
\begin{eqnarray}
u^\beta=u^{2z-2-2\nu+1}.
\end{eqnarray}
So $\psi_{\downarrow}$ being square integrable requires that $-\nu>-z$.
Combining the assumption that $z-\theta+d-1\geq 0$, the square integrability of $\psi_{\downarrow}$ results in $-z+d-1<\theta\leq z+d-1$.

For the case of $\nu$ vanishing, the leading behavior is the logarithmic term $\psi_{\uparrow}$ in Eq. (\ref{alpha1=alpha2}).
So, $u^{2z-2}\psi^2$ has the following behavior
\begin{eqnarray}
u^{2z-2}\psi^2=u^{2z-1}(\ln u)^2,
\end{eqnarray}
which is independent of $\theta$ and so square integrable near $u=0$ for any $z>1$.

Thus, for $0\leq\nu<z$, i.e., $-z+d-1<\theta\leq z+d-1$,
$\psi_{\downarrow}$ and $\psi_{\uparrow}$ are both in the Hilbert space near $u=0$.
While for $\nu\geq z$, only  $\psi_{\uparrow}$ is in the Hilbert space, without considering the eigenvalue $\lambda$.

Furthermore, as we discussed in Appendix B.1,  for $0\leq\nu<z$, the more general solution $\psi_2$ (\ref{psi2uinfty})
\begin{eqnarray}\label{psi2uinftyM}
\psi_2=C_{\downarrow}\psi_{\downarrow}+C_{\uparrow}\psi_{\uparrow},
\end{eqnarray}
which is fulfilled near the boundary  is square integrable under measure (\ref{Measure}).
Since the solutions multiplying the solution (\ref{psi2uinftyM}) by any phase also belong to the Hilbert space,
we have a one-dimensional solution space for each eigenvalue $\pm i$.
Therefore, a one-dimensional alternative set of
boundary conditions can be found in this range.
However, for $\nu\geq z$, because $\psi_{\downarrow}$ lies out of the Hilbert space,
so only one type of boundary condition is available.

\subsubsection{Boundary conditions}

Subsequently, we shall further study the available boundary conditions of the scalar field $\psi$ for $0\leq\nu<z$.
Similar to the discussion in the AdS and Lifshitz-AdS geometry\cite{1212.1728,gr-qc/0305012},
we define
\begin{eqnarray}
\label{Psigamma}
\psi_\gamma=\psi_{2,\lambda=i}+e^{i\gamma}\psi_{2,\lambda=-i}.
\end{eqnarray}
Considering the complete forms of $\psi_{\downarrow}$  and $\psi_{\uparrow}$ at $u=0$,
for $\nu$ being non-zero or a half integer, we simplify the behavior of $\psi_\gamma$ as
\begin{eqnarray}
\psi_\gamma\sim \sum_{n=0}^{\infty}(a_{n,\gamma} u^{\frac{1}{2}-\nu+n}+b_{n,\gamma}u^{\frac{1}{2}+\nu+n}),
\end{eqnarray}
where both $a_{n,\gamma}$ and $b_{n,\gamma}$ contribute to determine the choice of boundary conditions.
Although we can not match the solutions $\psi_{\uparrow,\downarrow}$ near $u=0$ to the solutions $\psi_{1,2}$ near $u=\infty$\footnote{Please see the Appendix B.},
as discussed in \cite{1212.1728}, we can still make a choice of extension,
which is determined by the ratio $b_{0,\gamma}/a_{0,\gamma}$.
Using Eq.(\ref{phiMomentum}), one can obtain the behavior of $\Psi$ near $u=0$
\begin{eqnarray}
\Psi_t\sim \sum_{n=0}^{\infty}(a_{n,\gamma} u^{\frac{z-\theta+d-1}{2}-\nu+n}+b_{n,\gamma}u^{\frac{z-\theta+d-1}{2}+\nu+n}).
\end{eqnarray}
The leading behavior is $u^{\frac{z-\theta+d-1}{2}-\nu}$. Actually, the exponent in the leading term is zero, which corresponds to the case with $m=0$ in AdS and Lifshitz-AdS geometry\cite{gr-qc/0305012,1212.1728}.

Finally, keeping the constrained conditions (\ref{NEC1}) and (\ref{NEC2}) in mind, we shall summarize our discussion about $\theta>0$ as follow:
\begin{itemize}
  \item When $\nu=0$, i.e., $z-\theta+d-1=0$, together with $z\geq 1$, which means $\theta\geq d$. This should be ruled out  because
    the range $\theta\geq d$ has no intersection with (\ref{NEC1}) or (\ref{NEC2}), and leads to instabilities of gravity.
  \item When $0<\nu<z$, i.e., $-z+d-1<\theta<z+d-1$, combining with the conditions (\ref{NEC1}) and (\ref{NEC2}), we have
\begin{eqnarray}
&&
\label{Rthetaz1}
1\leq z\leq 2,~~~~-z+d-1<\theta\leq (d-1)(z-1)~,
\
\\
&&
\label{Rthetaz2}
z>2,~~~~-z+d-1<\theta\leq d-1~.
\end{eqnarray}
In the parameter range above, the leading term of $\Psi_t$ is $a_{0,\gamma}u^{\frac{z-\theta+d-1}{2}-\nu}$ and the subleading term is
  $b_{0,\gamma}u^{\frac{z-\theta+d-1}{2}+\nu}$.
  Therefore, the choice of the ratio $\frac{b_{0,\gamma}}{a_{0,\gamma}}$ will correspond to
  different types of boundary conditions ranging from Dirichlet condition ($\frac{b_{0,\gamma}}{a_{0,\gamma}}=0$), Neumann condition ($\frac{b_{0,\gamma}}{a_{0,\gamma}}=\infty$) to Robin condition ($\frac{b_{0,\gamma}}{a_{0,\gamma}}$ is the value being neither 0 nor $\infty$).
  Note that since $\nu=\frac{z-\theta+d-1}{2}$, the exponent in the leading term is zero.
  It just corresponds to the case of $m^2=0$ in AdS or Lifshitz-AdS geometry\cite{gr-qc/0305012,1212.1728}.
  \item When $\nu\geq z$, we have $\theta\leq-z+d-1$, which falls into the range $\theta\leq(d-1)(z-1)$ (Eq.(\ref{NEC1}))
  or into $\theta\leq (d-1)$ (Eq.(\ref{NEC2})).
  For this case, the leading term is not available while only the subleading term is available,
  of which the form goes like $b_{0,\gamma}u^{\frac{z-\theta+d-1}{2}+\nu}$.
  Therefore, only one type of boundary condition is available in this case.
\end{itemize}

\subsection{Boundary conditions for $\theta<0$}
In this subsection, we turn to explore the possible boundary conditions for the case with $\theta<0$.
\subsubsection{Solution spaces $\mathcal{K}_{\pm}$}
Similar to the discussion in last subsection, we start with finding the solution of the eigenequation  at the singular point $u=0$.
From Eqs.(\ref{KGv2}) and (\ref{Qu}), it is explicit that $u=0$ is
an irregular point with rank $l=-\frac{\theta}{d-1}$.
This is very different from the observation for $\theta>0$, in which
$u=0$ is a regular singular point.
Note that with the same trick  in the above subsection, we can assume $\theta=-(d-1)l$ where $l=1,2,3,\ldots$.
If $\theta$ is a rational non-integer, a transformation of coordinates of $\rho=u^{1/n}$ can be made to fullfil
that $-\frac{n\theta}{d-1}$ is an integer.

Therefore, the solution of Eq.(\ref{KGv2}) has the form of Eq.(\ref{EFil})
\begin{eqnarray}\label{EFilinfinitypsiv2}
\psi(u)=F(u)\exp\left[\sum_{l=1}^{l}C_l u^{-l}\right],
\end{eqnarray}
with
\begin{eqnarray}\label{Fv2}
F(u)=\sum_{n=0}^{\infty}a_n u^{\alpha+n},~~~~~~a_0\neq 0.
\end{eqnarray}
Putting Eq.(\ref{EFilinfinitypsiv2}) back into Eq.(\ref{KGv2}), we have
\begin{eqnarray}
\label{Fv2}
\left[\sum_{l=1}^lC_ll(l+1)u^{-l-2}+\left(\sum_{l=1}^lC_l(-l)u^{-l-1}\right)^2+Q(u)\right]F
+\partial_u^2F+2\sum_{l=1}^lC_l(-l)u^{-l-1}\partial_uF=0,
\nonumber
\\
\end{eqnarray}
with $l=-\frac{\theta}{d-1}$.

It is easy to find that the most dominant terms at $u=0$ in Eq.(\ref{Fv2}),
i.e., the highest power term of $1/u$ are the squared sum term and the $m^2$ term.
The coefficient of the most dominant terms should be set to zero to satisfy the equation, so that we have
\begin{eqnarray}
\label{czv2}
C_\theta^2 \frac{\theta^2}{(d-1)^2}-m^2=0
\end{eqnarray}
which gives two roots $C_{\theta,\pm}$ as follow
\begin{eqnarray}
&&
C_{\theta,\pm}^{I}=\pm\left(-\frac{d-1}{\theta}\right)\sqrt{m^2},~~~~~~m^2\geq 0,
\
\\
&&
C_{\theta,\pm}^{II}=\pm\left(-\frac{d-1}{\theta}\right)\sqrt{|m|^2}\,i,~~~~m^2< 0.
\end{eqnarray}
Then to the leading order at $u=0$, the asymptotic behaviors of $\psi$ in Eq.(\ref{EFilinfinitypsiv2}) are
\begin{eqnarray}
\label{psi1psi2II}
&&
\psi_1^I=a_0u^{\alpha}\exp[C_{\theta,+}^Iu^{\frac{\theta}{d-1}}],~~\psi_2^I=a_0u^{\alpha}\exp[C_{\theta,-}^Iu^{\frac{\theta}{d-1}}]~~\mathrm{for}~~m^2\geq 0,
\
\\
&&
\psi_1^{II}=a_0u^{\alpha}\exp[C_{\theta,+}^{II}u^{\frac{\theta}{d-1}}],~~\psi_2^{II}=a_0u^{\alpha}\exp[C_{\theta,-}^{II}u^{\frac{\theta}{d-1}}]~~\mathrm{for}~~m^2< 0.
\end{eqnarray}
Since Eq.(\ref{czv2}) is independent of the eigenvalue, the above solutions satisfy the eigenequation $A\psi=\pm i \psi$.
In addition, the above solutions are square integrable except $\psi_1^I$.

Moreover, in Appendix B.2, we discussed that in this case, the general possible square integrable solution near the boundary is
$\psi_2$ in Eq. (\ref{psi2theta-}), i.e.,
\begin{eqnarray}\label{psi2theta-M}
\psi_2=C_1\psi_1^{j}+C_2\psi_2^j.
\end{eqnarray}
Because the exponent damped solution $\psi_1^I$  lives beyond of the Hilbert space and we can not construct the above
linear combination for $j=I$, so only one type of boundary conditions is available for the modes with $m^2\geq 0$.
But for $m^2<0$,  the construction in Eq.(\ref{psi2theta-M}) is available in Hilbert space and any solution multiplying
that is included in, so there is a one-dimensional alternative set of boundary conditions.

\subsubsection{Boundary conditions}

Following the steps  in the case with $\theta>0$, we can deduce the behavior of $\psi_{\gamma}$ for $\theta<0$ and $m^2<0$ as
\begin{eqnarray}
\psi_\gamma\sim \sum_{n=0}^{\infty}\left(\tilde{a}_{n,\gamma}u^{n} \exp[C_{\theta,+}^{II}u^{\frac{\theta}{d-1}}]
+\tilde{b}_{n,\gamma} u^{n} \exp[C_{\theta,-}^{II}u^{\frac{\theta}{d-1}}]\right),
\end{eqnarray}
where both $\tilde{a}_{n,\gamma}$ and $\tilde{b}_{n,\gamma}$ will determine the choice of boundary conditions. Furthermore,
 the behavior of $\Psi$ near $u=0$ can be written as
\begin{eqnarray}
\Psi_t\sim \sum_{n=0}^{\infty}\left(\tilde{a}_{n,\gamma}u^{\frac{z-\theta+d-2}{2}+n} \exp[C_{\theta,+}^{II}u^{\frac{\theta}{d-1}}]
+\tilde{b}_{n,\gamma} u^{\frac{z-\theta+d-2}{2}+n} \exp[C_{\theta,-}^{II}u^{\frac{\theta}{d-1}}]\right),
\end{eqnarray}
which can be expanded into the form
\begin{eqnarray}
\Psi_t\sim \sum_{n=0}^{\infty}\left(\tilde{a}_{n,\gamma}u^{\frac{z-\theta+d-2}{2}+n}[1+C_{\theta,+}^{II}u^{\frac{\theta}{d-1}}+\ldots]
+\tilde{b}_{n,\gamma} u^{\frac{z-\theta+d-2}{2}+n}[1+C_{\theta,-}^{II}u^{\frac{\theta}{d-1}}+\ldots]\right).
\end{eqnarray}

Therefore, we shall close this subsection with a summary of the possible
boundary conditions of the scalar field in the HV geometry with $\theta<0$. For the modes with $m^2<0$, there are different types of boundary conditions
ranging from Dirichlet condition, Neumann condition to Robin condition, which
is set by the ration $\frac{\tilde{b}_{0,\gamma}}{\tilde{a}_{0,\gamma}}$.
While for the modes with $m^2\geq 0$, only one type of boundary conditions is available.

\section{Conclusions and discussions}\label{Conclusions}

In this letter, we examined the normalizablity and studied the possible boundary conditions of scalar field in hyperscaling violating geometry.
We discussed them in both cases with $\theta>0$ and $\theta<0$, respectively. In the case with $\theta>0$,  we fixed two sets of range of geometrical
parameters, HV exponent $\theta$ and Lifshitz exponent $z$, in which the types of possible boundary conditions can be different. Specifically,
in a certain range, see Eqs. (\ref{Rthetaz1}) and (\ref{Rthetaz2})),
three types of boundary conditions including Dirichlet condition, Neumann condition to Robin
condition, are available for the scalar mode.  This observations is analogous to the well known mass window in AdS geometry  allowing for different quatizations of the scalar modes. However,
the difference is that, in this case,
the mass of  the scalar field does not play any role in determining the possible boundary conditions in HV geometry.
While in the range of $\theta\leq-z+d-1$, i.e., $\nu\geq z$, only the subleading term of $\Psi_t$ survives,
which leads to the conclusion that no alternative choice of the boundary conditions is available and only Dirichlet type condition can be set.
Also, we studied them parallelly in the case with $\theta<0$, in which
we found that for $m^2<0$, three types of boundary conditions can be set while for
$m^2>0$, only one type is available.

Here we only consider the scalar field in the HV geometry but never refer to any specific model.
It would be very interesting to explore the scalar, vector and tensor fluctuations on specific HV models,
such as Refs. \cite{1112.0573,1201.1905,1112.2702,1209.3946} and references therein.
However, we would like to point out that considering the specification of a HV background involves a specific metric and a running dilaton,
 it is still not obvious that the KG mass is an interesting coupling in the context of holography for HV backgrounds.
The mass that determines the dimensions of the dual scalar operators is the mass in the dual frame, which can be obtained by Weyl transformation \cite{KanitscheiderKD,DuffFG}.
This mass term in the dual frame may be more relevant for the physics of the dual theory, which calls for further understanding.

\begin{appendix}\label{Appendix}

\section{Series solutions of the second order homogeneous linear differential equations}

In this appendix, we will give a brief summary of the series solutions of the second order homogeneous linear differential equations.
For the detailed discussions, please refer to\cite{Arfken,Bender,MIT}.
Now, we begin with the following differential equation
\begin{eqnarray}\label{LEv1}
y''(x)+P(x)y'(x)+Q(x)y(x)=0.
\end{eqnarray}
Usually, a point $x_0$ can be classified as the ordinary point, regular singular point and irregular singular point.
The ordinary point is that the coefficient functions $P(x)$ and $Q(x)$ are all analytic at point $x_0$
\footnote{Here, we assume that $x_0$ is finite. For an infinite point, one can set $x=1/\chi$ and then study the behavior as $\chi\rightarrow 0$.}.
For ordinary point, $y(x)$ can be expanded in term of Taylor series
\begin{eqnarray}\label{Taylor}
y(x)=\sum_{n=0}^{\infty}c_n(x-x_0)^n.
\end{eqnarray}

A singular point is that at least one of the coefficient functions $P(x)$ and $Q(x)$ are not analytic at $x_0$.
The singular point $x_0$ can be furthermore classified as regular singular point and irregular singular point.
If all of $(x-x_0)^2 Q(x), (x-x_0)P(x)$ are analytic at $x=x_0$, the point $x_0$ is regular singular point.
Otherwise, the point $x_0$ is called irregular singular point. Especially, if
\begin{eqnarray}\label{rankl}
(x-x_0)^{l+1}P(x)=\sum_{n=0}^{\infty}a_n(x-x_0)^n,~~~~~(x-x_0)^{2(l+1)}Q(x)=\sum_{n=0}^{\infty}b_n(x-x_0)^n,
\end{eqnarray}
where at least one of $a_0$ and $b_0$ are not zero and $l$ is a positive integer, then the point $x_0$ is the irregular singular point of rank $l$.
If the order of the pole of $P(x)$ and that of $Q(x)$ at $x_0$ are $l_1+1$ and $2l_2+2$ respectively, with $l_1$ not equal to $l_2$,
then the rank $l$ is equal to the greater of $l_1$ and $l_2$.

For a regular singular point, instead of the Taylor series, one needs a Frobenius series
\begin{eqnarray}\label{Frobenius}
y(x)=(x-x_0)^\alpha\sum_{n=0}^{\infty}c_n(x-x_0)^n,~~~~~~c_0\neq 0.
\end{eqnarray}
Substituting the above expansion into the second order differential equation (\ref{LEv1}),
and then requiring that the coefficient for each power of $x-x_0$ must vanish separately,
one can obtain the following indicial equation
\begin{eqnarray}\label{Indicial}
\alpha(\alpha-1)+p_0\alpha+q_0=0,
\end{eqnarray}
where $p_0=\lim_{x\rightarrow x_0}(x-x_0)P(x)$ and $q_0=\lim_{x\rightarrow x_0}(x-x_0)^2Q(x)$.
The above equation gives two values $\alpha_1$, $\alpha_2$.
If $\alpha_1-\alpha_2\neq j,~j=0,1,2,\ldots$, there are two linearly independent solutions of Frobenius form
\begin{eqnarray}\label{Frobeniusv11}
&&
y_1(x)=(x-x_0)^{\alpha_1}\sum_{n=0}^{\infty}a_n(x-x_0)^n,~~~~~~a_0\neq 0,
\
\\
\label{Frobeniusv12}
&&
y_2(x)=(x-x_0)^{\alpha_2}\sum_{n=0}^{\infty}b_n(x-x_0)^n,~~~~~~b_0\neq 0.
\end{eqnarray}
If $\alpha_1-\alpha_2=0$, one of the two linearly independent solution, $y_1(x)$, is the Frobenius form (\ref{Frobeniusv11}) and another,
$y_2(x)$, looks like
\begin{eqnarray}\label{Frobeniusv2}
y_2(x)=y_1(x)\ln(x-x_0)
+(x-x_0)^{\alpha_2}\sum_{n=0}^{\infty}b_n(x-x_0)^n.
\end{eqnarray}
If $\alpha_1-\alpha_2=j$, with $j=1,2,\ldots$, $y_1(x)$ still remains the form (\ref{Frobeniusv11}),  but $y_2(x)$ will be
\begin{eqnarray}\label{Frobeniusv3}
y_2(x)=Cy_1(x)\ln(x-x_0)
+(x-x_0)^{\alpha_2}\sum_{n=0}^{\infty}b_n(x-x_0)^n,~~~~~~b_0\neq 0,
\end{eqnarray}
where $C$ is a constant that might vanish or not.

For the irregular singular point of rank $l$, the solutions are of the form
\begin{eqnarray}\label{EFil}
y(x)=F(x)\exp\left[\sum_{l=1}^{l}C_l(x-x_0)^{-l}\right],
\end{eqnarray}
where $F(x)$ is a Frobenius series.

If the irregular singular point of rank $l$ is infinity, then the solutions of Eq.(\ref{LEv1}) at large $x$ is of the form
\begin{eqnarray}\label{EFilinfinity}
y(x)=\left(\sum_{n=0}^{\infty}a_n x^{-\alpha-n}\right)\exp\left[\sum_{l=1}^{l}C_lx^{l}\right],~~~~~~a_0\neq 0.
\end{eqnarray}

\section{The square integrability at infinity}
Since Eq. (\ref{KGv2}) has two singular points, i.e., $u=0$ and $u\rightarrow \infty$, for completeness, in this Appendix,
we shall also present the discussions on the possible square integrability of solutions to Eq. (\ref{KGv2}) under the measure (\ref{InnerProductMomentum}).

\subsection{Case I: $\theta>0$}
To search for the possible square integrable solutions to Eq.(\ref{KGv2}) over the full radial range,
we need to further exploit the behavior of these solutions at infinity.
To deal with the behaviors at infinite point, one usually sets $u=1/r$ and study them at $r\rightarrow 0$.
Under this transformation,  the eigenequation (\ref{KGv2}) becomes
\begin{eqnarray}
\label{KGr}
\partial_r^2\psi+\frac{2}{r}\partial_r\psi+\tilde{Q}(r)\psi=0,
\end{eqnarray}
where $\tilde{Q}$ has the form
\begin{eqnarray}
\label{tildeQr}
\tilde{Q}(r)=-p(p+1)r^{-2}-m^2r^{-2(1+\frac{\theta}{d-1})}-k^2r^{-4}+\lambda r^{-2(1+z)}.
\end{eqnarray}
From Eq.(\ref{KGr}), we can see that the singular point $u=\infty$ ($r=0$) is an irregular singular  point
with rank $z$  if $z\geq\frac{\theta}{d-1}$ and with rank $\frac{\theta}{d-1}$ if $z<\frac{\theta}{d-1}$, respectively.
Note that $z<\frac{\theta}{d-1}$ is excluded due to the gravitational instability, then we can only focus on the case with $z\geq\frac{\theta}{d-1}$
and also we will assume integer $z=l, l=2,3,4,\ldots$
\footnote{For rational and noninteger $z$, we can make a transformation of  coordinate as $\rho=r^{1/n}$ so that we have the rank $nz$ is an integer.}.

For an irregular singular point with rank $l$ at infinity, the solution of Eq.(\ref{KGv2}) has the form of (\ref{EFilinfinity}), i.e.,
\begin{eqnarray}\label{EFilinfinitypsi}
\psi(u)=F(u)\exp\left[\sum_{l=1}^{l}C_l u^{l}\right]~~~~\mathrm{with}~~~~F(u)=\sum_{n=0}^{\infty}a_n u^{-\alpha-n},~~~~~~a_0\neq 0.
\end{eqnarray}

Substituting Eq.(\ref{EFilinfinitypsi}) into Eq.(\ref{KGv2}), we obtain
\begin{eqnarray}
\label{F}
&&
\left[\sum_{l=1}^lC_ll(l-1)u^{l-2}+\left(\sum_{l=1}^lC_llu^{l-1}\right)^2-p(p+1)u^{-2}-m^2u^{\frac{2\theta}{d-1}-2}-k^2+\lambda u^{2z-2}\right]F
\nonumber
\\
&&
+\partial_u^2F+2\sum_{l=1}^lC_llu^{l-1}\partial_uF=0,
\end{eqnarray}
where $l=z$. We pick up the largest power terms in the left hand side of Eq.(\ref{F}) and require that its coefficients vanish,
such that we get the following equation
\begin{eqnarray}
\label{cz}
C_z^2 z^2+\lambda=0.
\end{eqnarray}
The above equation has no print of $\theta$ and only depends on the Lifshitz exponent $z$,
which agrees well with that found in Lishitz geometry\cite{1212.1728}.
Eq.(\ref{cz}) gives  two roots $C_{z,\pm}$, so that we have  the asymptotic behaviors of the scalar field to the leading order near $u=\infty$ as
\begin{eqnarray}
\label{psi1psi2}
\psi_1=a_0u^{-\alpha}\exp[C_{z,+}u^z],~~~\psi_2=a_0u^{-\alpha}\exp[C_{z,-}u^z].
\end{eqnarray}

Then, we have to fix $C_{z,\pm}$ by  the chosen eigenvalue $\lambda=\pm i$
with which the eigenequation
gives a square integrable solution space $\mathcal{K}_{\pm}$ under the measure $V^{-1}d\Sigma$\cite{gr-qc/0305012}.
Due to $a_0\neq 0$, Eq.(\ref{cz}) has roots
\begin{eqnarray}
&&
\label{Cz1}C_{z,\pm}^{i}=\pm\left(\frac{1+i}{\sqrt{2}z}\right),~~\mathrm{for}~~\lambda=i,
\\
&&
\label{Cz2}C_{z,\pm}^{-i}=\pm\left(\frac{1-i}{\sqrt{2}z}\right),~~\mathrm{for}~~\lambda=-i.
\end{eqnarray}
Since near the infinity $u=\infty$, the exponential function in Eq.(\ref{psi1psi2}) blows up exponentially for $C_z=C_{z,+}$ while exponentially damped for $C_z=C_{z,-}$,  the solution $\psi_2$ is alway square integrable while $\psi_1$ is not  at the singular point $u=\infty$.
Furthermore, we can express the solution $\psi_2$ at $u=\infty$ in term of  the linear combination of those at $u=0$ as
\begin{eqnarray}
\label{psi2uinfty}
\psi_2=C_{\downarrow}\psi_{\downarrow}+C_{\uparrow}\psi_{\uparrow},
\end{eqnarray}
where $C_{\downarrow}$ and $C_{\uparrow}$ are constant.
Therefore, over the full range from $u=0$ to $u=\infty$, the possible solutions of square integrability under measure (\ref{Measure})
are $\psi_2$.
Taking account of the behaviors of $\psi$ at $u=0$ and that at infinity, we can conclude that
for $0\leq\nu<z$, in the range from $u=0$ to $u=\infty$, the solution $\psi_2$ is square integrable under measure (\ref{Measure}).

\subsection{Case I: $\theta<0$}
Now we shall discuss the behaviors of KG equation at infinity for $\theta<0$.
From Eq.(\ref{KGr}), we can easily find that the point at infinity ($u=\infty$)
is an irregular singular point with rank $z$.
Therefore, the behavior of eigenequation (\ref{KGv2}) for $\theta<0$ at infinity is the same as that for $\theta>0$,
i.e., Eq.(\ref{psi1psi2})-Eq.({\ref{Cz2}}), which are independent of $\theta$ and only dependent of $z$.

Obviously, the square integrable solutions $\psi_2$  at infinity (see Eq.(\ref{psi1psi2}) ) for $\lambda=\pm i$
can be expressed as the linear combination of $\psi_1^{j}$ and $\psi_2^{j}$ at $u=0$,
\begin{eqnarray}\label{psi2theta-}
\psi_2=C_1\psi_1^{j}+C_2\psi_2^j
\end{eqnarray}
with constant $C_1$, $C_2$ and $j=I,II$. Therefore, in the range from $u=0$ to $u=\infty$, the solution
$\psi_2$ is possible square integrable under measure (\ref{Measure}).

\end{appendix}

\begin{acknowledgments}
We are grateful to the anonymous referees for valuable suggestions and comments, which
are important in improving our work.
J. P. Wu is supported by the Natural Science Foundation of China under Grant
Nos. 11305018 and 11275208 and also supported by Program for Liaoning Excellent Talents in University (No. LJQ2014123). X. M. Kuang is funded by FONDECYT grant No.3150006
and the PUCV-DI Projects No.123.736/2015.
\end{acknowledgments}


\end{document}